\documentclass[12pt]{iopart}

\usepackage{iopams}  
\usepackage{comment}

\usepackage{caption,setspace}
\usepackage[lofdepth,lotdepth]{subfig}
\usepackage{tikz}

\usepackage{pgfplots}
\usepackage{pgfplotstable}
\usetikzlibrary{calc,arrows,patterns, fadings,shapes,calc,shadows,svg.path}
\pgfplotsset{compat=newest}
\usetikzlibrary{external}

\begin{document}

\title[IOP Publishing journals]{Inverse design of waveguide mode converters using artificial neural networks}

\author{Ali Mohajer Hejazi$^1$, Vincent Ginis$^1$$^,$$^2$$^,$$^3$}

\address{$^1$Applied Physics, Vrije Universiteit Brussel, 1050 Brussel, Belgium}
\address{$^2$Data Lab, Vrije
Universiteit Brussel, 1050 Brussel, Belgium}
\address{$^3$Harvard John A. Paulson School of Engineering and Applied Sciences, Harvard University, Cambridge, MA 02138, USA}
\ead{ali.mohajer.hejazi@vub.be}
\vspace{10pt}
\begin{indented}
\item[]
\end{indented}

\begin{abstract}
Machine learning techniques, notably various deep neural network methods, are instrumental in processing extensive and intricate data sets in engineering and scientific fields. This paper shows how deep neural networks can inversely design cascaded-mode converting systems, particularly the waveguide gratings that implement selective mode conversion upon reflection. Neural networks can map the grating's physical features to scattering parameters of the modes reflected from the grating. The trained networks can then be utilized to inversely design the gratings based on the desired values of the scattering parameters. The process of the inverse design involves using the technique of gradient descent of a defined loss function. Minimizing this loss function leads to calculating more accurate features fulfilling the desired scattering parameters.

\end{abstract}

%
%
%
%
\ioptwocol

\section{Introduction}
The performance of conventional photonic components, such as waveguides, gratings, and plasmonic structures, depends on the size of their geometrical features relative to the wavelength of light. These components may be designed using pure analytical models or based on prior physical intuitions~\cite{jiang2021deep, mao2021inverse, molesky2018inverse}. In many photonic design challenges, such as those involving a wide operational bandwidth, nonlinear phenomena, or the need for high-density integration, our understanding may need to deepen to design the most efficient devices due to the interdependence of multiple physical phenomena co-occurring. Evaluating the performance of light interaction with these photonic components requires using numerical techniques in electromagnetism, which may not be intuitive when the geometrical features and constitutive materials of the device are complex~\cite{so2020deep, molesky2018inverse}.
In conjunction with the conventional methods mentioned above, the inverse design approach can be employed to enhance accuracy and efficiency in the design process of photonic components. Forward design problems involve analyzing electromagnetic components through numerical simulations or mathematical tools to determine their responses when interacting with electromagnetic waves. Examples of these include the analysis of scattering parameters, radiation patterns, and transmission coefficients.

Inverse problems involve reversing the forward problem, meaning that if we aim to achieve a specific response from a device, we must systematically modify the device's geometrical and material properties to obtain that response~\cite{jiang2021deep}. Finding solutions for inverse problems is challenging due to the non-convexity of the solution space, which arises from numerous local minima~\cite{campbell2019review}.
Generally, inverse problems fall under the broader field of optimization problems, requiring the application of optimization techniques to solve them. Optimization methods prevalent in the photonics community include evolutionary algorithms~\cite{wiecha2019design}, such as genetic algorithms~\cite{haupt2007genetic, shi2017optimization, jafar2018adaptive} and particle swarm algorithms~\cite{mikki2008particle, li2019inverse, chung2020tunable}, topology optimization~\cite{molesky2018inverse, jensen2011topology, christiansen2021inverse}, and optimization aided by deep neural networks, among others.
Over the past decade, many researchers in the photonics community have focused on using neural networks for both forward and inverse problems. Deep neural networks excel at finding complex nonlinear mappings between the topological and material features of photonic components and their related functional properties~\cite{so2020deep}. Two clear examples of deep neural networks' utility in nanophotonics are presented in both Purifoy et al. and Lenaerts et al.'s research articles~\cite{peurifoy2018nanophotonic, lenaerts2021artificial}. In the latter study, a thin dielectric material layer serves as a Fabry–Perot resonator. When broadband light enters the resonator, numerous partial reflections emerge from the resonator boundaries due to the differences in refractive indices inside and outside the resonator. The interference between these partial waves significantly impacts the resonator's reflection and transmission spectra. This interference mechanism depends on the resonator's physical features, such as dimensions and dielectric constant, meaning that altering these features determines the spectral response of the resonator. In both studies, the authors collected training data and designed a deep neural network to map the input—nodes representing physical features—to the output, the transmission spectrum with 200 nodes. Once training was complete, the trained model functioned as a fixed function on which gradient descent was calculated to obtain the physical characteristics that would achieve the desired transmission spectrum.

Many different types of neural networks have proven effective in solving various photonics problems. For example, convolutional neural networks (CNNs) have demonstrated their utility in extracting key geometrical features from complex-shaped nanostructures~\cite{sajedian2019finding, an2022deep}. Another example is generative adversarial networks (GANs), which consist of two networks: a generator and a discriminator. The generator creates structural images from noise, which are expected to possess desired optical properties, while the discriminator evaluates the generator's performance. The generator aims to deceive the discriminator by producing fake images until the discriminator can no longer distinguish between real and fake images~\cite{liu2018generative, jiang2019free, so2019designing}.

In this work, we want to provide an easy-to-understand tutorial of how integrated waveguide gratings can be designed using inverse design based on deep neural networks. Integrated waveguide gratings are found in numerous photonic applications, such as wavelength filters~\cite{venghaus2006wavelength}, modal conversion~\cite{yariv1977periodic}, remote manipulation of near-fields~\cite{ginis2020remote}, distributed feedback lasers, and distributed Bragg reflector lasers~\cite{carroll1998distributed}, among others. In this paper, we focus on the type of gratings used in cascaded-mode conversion~\cite{ginis2020remote,ginis2023resonators}.

\section{A grating waveguide as a mode converter}
\label{section:grating_waveguide_simulation}


We begin by designing a straight, two-dimensional dielectric waveguide capable of guiding the first three TE modes at a wavelength of $1550 nm$, as illustrated in Fig.~1. The dielectric material used in this waveguide structure is silicon, with a refractive index of 3.48. Figures~1(b-d) illustrate the electric field distributions of the first three TE modes of the waveguide in the absence of the grating. The waveguide is positioned at the center, and the evanescent electric fields of the guided modes are plotted near the dielectric waveguide's boundary. According to optical waveguide theory, all guided modes of an unperturbed dielectric waveguide are orthogonal to each other~\cite{alferness2013guided}. The overlap integral of the electric fields of orthogonal modes over the transverse waveguide profile thus equals zero. When a perturbation, such as a grating structure, is introduced in the waveguide's longitudinal direction, the orthogonality condition is violated, leading to inevitable energy exchange between the modes. In essence, the grating acts as a distributed radiating source capable of feeding energy into or out of the waveguide modes~\cite{yariv2007photonics}. For our purpose, there are two types of grating structures: one with longitudinal corrugations created by periodically removing the dielectric portion of the waveguide, and another comprising a periodic arrangement of two different dielectric materials in terms of their refractive indices, stacked together in the longitudinal direction. In this paper, we focus on the corrugated structure (shown in Fig.~1(a)). The primary condition required for a grating structure to convert two modes of the same type is known as longitudinal phase matching~\cite{yariv1977periodic}, which is given by:

\begin{equation}
\beta_\mu + \beta_\nu - m(\frac{2\pi}{\Lambda})=0
\label{formula:phase_matching_condition}
\end{equation}

In this formula, \(\beta_{\mu}\) is the propagation constant of the excited mode at the beginning of the waveguide, \(\beta_{\nu}\) is propagation constant of the desired mode to be converted, \(\Lambda\) is the period of the grating, and the parameter \textit{m} is an arbitrary integer. In the formula (\ref{formula:phase_matching_condition}), both waves are contra-directional, meaning that the converted wave is reflected back from the grating.

\begin{figure*}[h!]
\centering
\includegraphics[width=0.9\textwidth]{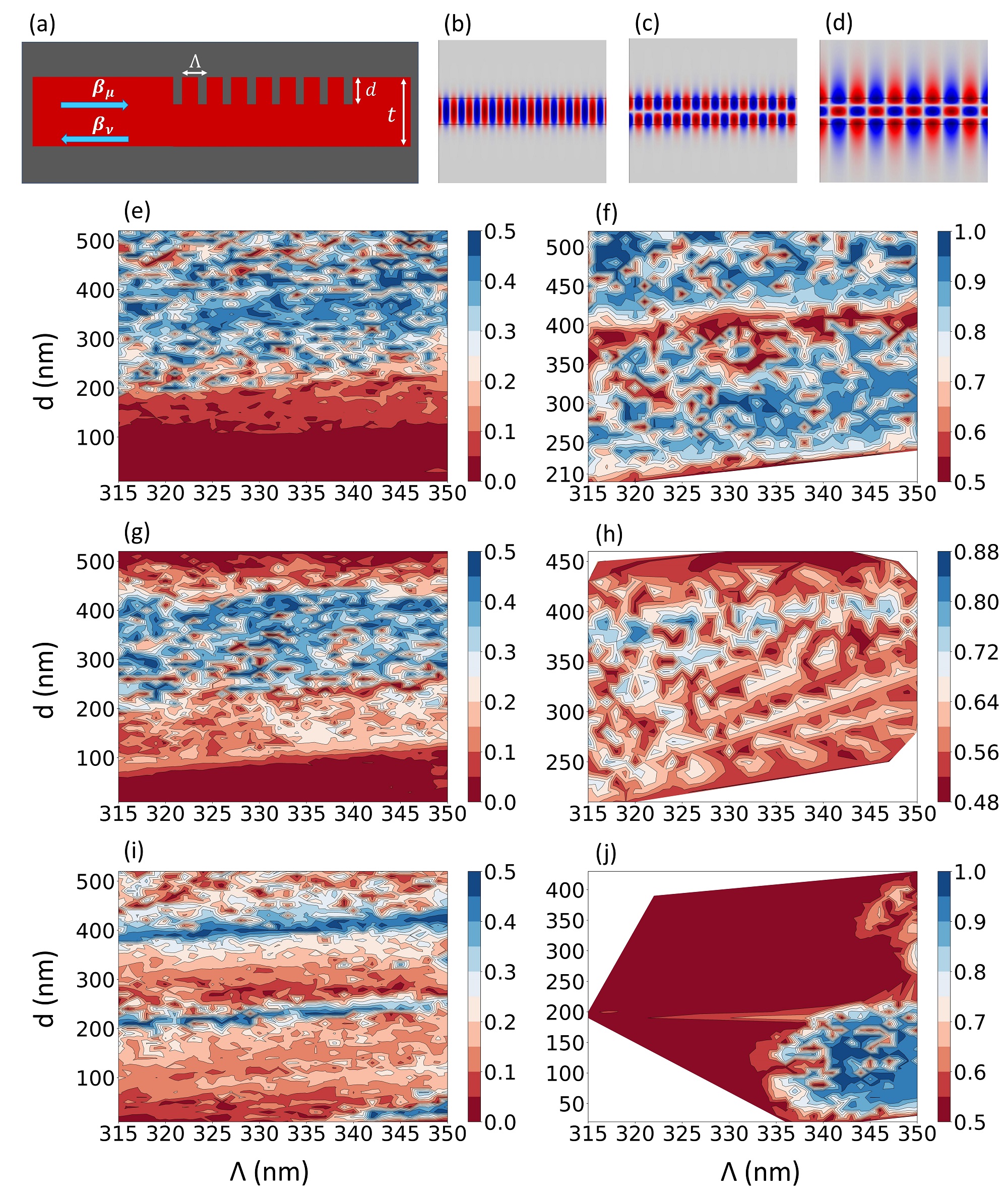}
\caption{(a) illustrates the waveguide and the grating structure. The parameters \(\Lambda\), $\mathbf{t}$, and $\mathbf{d}$ are the grating period, the waveguide thickness and the grating corrugation depth respectively. The chosen value for the parameter $\mathbf{t}$ is equal to 525 nm. The deep grey area around the waveguide is vacuum. (b)-(d) illustrate the plots of the electric field distributions of the first three TE modes of the dielectric waveguide. (e),(g) and (i) depict the contour plots of the ground-truth values of $|S_{11}|$$<$ 0.5, $|S_{21}|$$<$ 0.5, and $|S_{31}|$$<$ 0.5 respectively. The plots are showing the variations of the scattering parameters over the \(\Lambda\) and $\mathbf{d}$. (f),(h) and (j) are similar to the plots in figures (e),(g) and (i), however the scattering parameters are shown for the magnitude larger than 0.5.}
\label{figures1}
\end{figure*}

We use a Finite Elements software COMSOL Multiphysics for simulation of the waveguide grating. In figure~1(a), the general appearance of the waveguide grating is illustrated. We alter physical features of the grating structure i.e. period, depth of the corrugations and duty cycle. We investigate behavior of the grating when one specific mode is excited. The parameters obtained from the simulation are scattering parameters of different modes interacting with the grating. For instance, if the first TE mode of the waveguide is excited, the achieved parameters are $S_{11}$, $S_{21}$ and $S_{31}$ indicating the reflection of the first, the second and the third mode from the grating respectively. The parameters $S_{21}$ and $S_{31}$ represent how much of the signal carried by the first mode is converted to the second and the third mode respectively. The simulations for obtaining the scattering parameters are conducted at a fixed optical wavelength of 1550 nm, however, the physical features of the grating i.e. period, corrugation depth and, duty cycle, are swept in the range of 315 to 350 nm, 10 to 520 nm and, 10 to 90 percents respectively. Figures~1(e-j) illustrate contour plots of the parameters $|S_{11}|$, $|S_{21}|$ and, $|S_{31}|$ as functions of the period and the corrugation depth. The sign $|.|$ represents the absolute value of the mentioned scattering parameters. The figures~1(e),(g),(i) are referred to parameters $|S_{11}|$, $|S_{21}|$ and, $|S_{31}|$ in the range between 0 and 0.5. Also, figures~1(f),(h),(j) illustrate contour plots of absolute values of the scattering parameters in the range between 0.5 and 1. From figure~1(f), it is fairly clear that a wide part of the surface of the contour plot indicates the possibility of obtaining large value of $|S_{11}|$ by choosing right values for the period and the corrugation depth. The trick here is choosing a proper value for the duty cycle. For the scattering parameter $|S_{21}|$ plotted in figure~1(h), the regions possessing capabilities of giving high value of $|S_{21}|$ are not so wide. At values of the corrugation depth between 350-400 nm and many values of the period, it is possible to achieve large value of $|S_{21}|$. It is noteworthy again that another determining parameter is the duty cycle. In the figure~1(j), it is sort of straightforward to locate which region could possibly indicate high value of $|S_{31}|$. The region with the period between 340-350 nm, and the corrugation depth between 10-200 nm has the property of large $|S_{31}|$. 

\begin{figure}[h]
\centering
\includegraphics[scale=0.05]{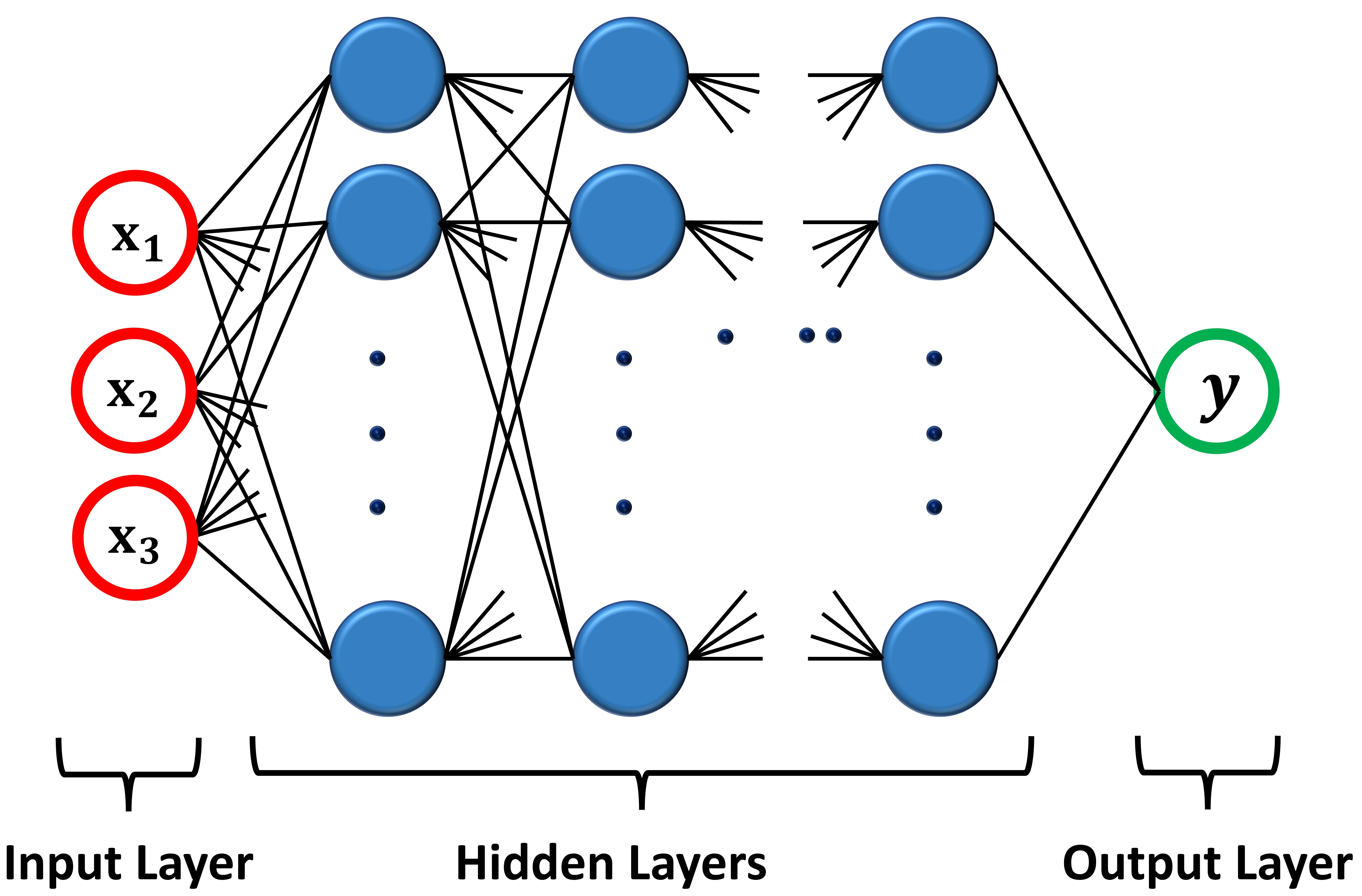}
\label{neural_network_structure}
\caption{Our neural network consists of input layer with 3 nodes, 5 hidden layers each with 600 nodes, and output layer with 1 node. In the input layer, the 3 nodes are physical features of the grating that are the period, the corrugation depth and the duty cycle respectively. For each hidden layer, a dropout layer is used for preventing the neural network from overfitting.}
\end{figure}

\section{Neural network modeling of the grating waveguide}
\label{section:grating_waveguide_NN}
In the section~\ref{section:grating_waveguide_simulation}, we discuss the data obtained from the electromagnetic simulations of the grating wavegide via a finite elements software. In the simulations, we alter the geometrical features of the grating to obtain scattering parameters. In this section, a neural network is trained to map the geometrical features of the grating as the network inputs to the output that is one of the scattering parameters $|S_{11}|$, $|S_{21}|$, or $|S_{31}|$. A number of hidden layers composes the space between the input and output layers as it is depicted in the figure~2. Each node in the layers represents a linear function with an alterable weight and a bias, and a nonlinear function known as activation. The output of the linear function plays the role of the input of the activation function.

\begin{figure*}
\centering
\includegraphics[width=0.9\textwidth]{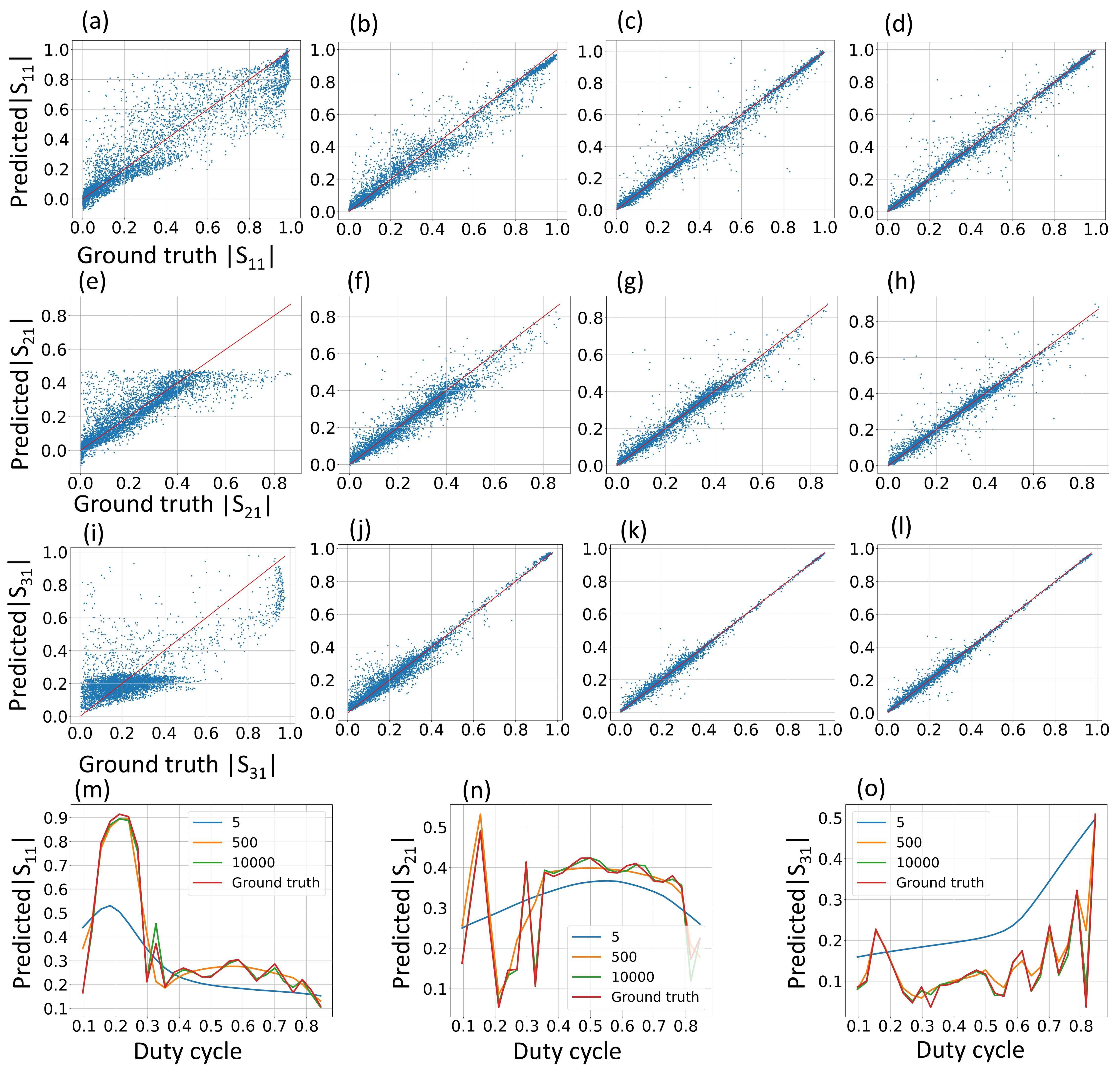}
\caption{(a)-(d) illustrate prediction-capability of the neural network designed to predict the scattering parameter $|S_{11}|$ when the epoch number is increased. The epoch number for the plots are chosen to be 5, 500, 5000, 10000 respectively. Each plot shows the neural network prediction (vertical axis) versus the ground-truth values of $|S_{11}|$ for a test set. The red line represents an ideal situation where the neural network has no error. Being closer to the red line means smaller error and consequently better prediction model for the data set. (e)-(h) and (i)-(l) are showing the same concept for $|S_{21}|$ and $|S_{31}|$ respectively. (m)-(o) depict the effect of increasing in epoch number on the model predictability in the case that the alteration of the scattering parameters versus duty cycle is investigated. The values of the corrugation depth and the period are chosen to be 270 and 348 (nm) respectively.}
\label{epochs_increasing_test}
\end{figure*}

In the hidden layers, each layer includes 600 nodes with a dropout layer. The latter layer is used for preventing the neural network from overfitting. The dropout layers randomly select nodes and deactivate them in the forward pass and in the process of backward propagation. Therefore, the weights corresponding to the dropped nodes are not updated~\cite{Dropout_layer}. The activation function used in the $1^{st}$, $2^{nd}$, $4^{th}$ and $5^{th}$ layers is Rectified-Linear-Unit function abbreviated as ReLU function. After conducting many experiments on different neural network architectures, we concluded that the activation function for the $3^{rd}$ hidden layer should be Sigmoid function. In the training process, the inputs pass through the network---the weights and the biases are initially randomly chosen---, then the calculated output result is compared with actual relevant scattering parameter via a chosen loss function. The gradient of this loss function is taken and is back propagated in the network to update the weights and the biases. The loss function that we used in our neural networks is logarithm of cosine hyperbolic of the prediction error which is given by:
\begin{equation}
\ L(y_{pred},y_{actual}) = \sum_{i=1}^{n} { \log }({ \cosh }(y^i_{pred}-y^i_{actual}))
\label{formula:loss_function}
\end{equation}

In the loss function formula (\ref{formula:loss_function}), $y_{pred}$ represents the absolute value of the scattering parameter predicted by the neural network and $y_{actual}$ is the actual value of the scattering parameter calculated from the simulation results. The whole data set passes the neural network in many epochs---each time that the whole data set passes the network back and forth, is known as one epoch---and the loss function diminishes gradually, so that the difference between the predicted value and the actual value decreases. In the figure \ref{epochs_increasing_test}, the effect of increase in epochs on prediction capability of our neural network is illustrated. Figure~3 illustrates predictability performance of three neural network models for $|S_{11}|$, $|S_{21}|$, and $|S_{31}|$. In each subfigure, the blue points show the scatter plot of the prediction---vertical axis---versus the actual values of the relevant scattering parameter. Increasing the epoch number, the blue points move closer to the red line illustrating the ideal situation that a neural network model might achieve. The blue points are actually the test data set that are applied to the trained neural network.
It is also worthwhile to note the effect of increase in epoch number when the values of the corrugation depth and the period are fixed but the duty cycle is changing. Figures~3(m-o) are results of three fairly different neural network models whose outputs are $|S_{11}|$, $|S_{21}|$, and $|S_{31}|$ respectively. Each figure shows the change of the scattering parameters over the duty cycle when the relevant neural network is trained for different epochs. It is evident from the figures that the increase in the epoch number would result in a closer network prediction to the actual data. In the section \ref{section:Inverse_design_grating_waveguide_NN} in addition to inverse design of the grating, some quantitative measures to evaluate the performance of the designed neural networks is discussed.

\section{Inverse design assisted by the neural network model}
\label{section:Inverse_design_grating_waveguide_NN}
In the section ~\ref{section:grating_waveguide_NN}, We discuss the neural network models capable of mapping the inputs---grating period, corrugation depth and the duty cycle---to the output that is absolute value of one of the scattering parameters. In the present section, we will discuss inverse design of the grating structure via the trained neural networks obtained in the previous section. This exactly means that we have a desired fixed value of one of the scattering parameters, and we want to achieve the physical features that could fulfill the desired value. The inverse design process is conducted through implementing the gradient descent algorithm from the output toward the input of the neural network. It is noteworthy that the weights and the biases of the trained neural network are fixed in the process. The inverse design algorithm is initialized by a random selection of the values of the physical features of the grating and then we apply them to the trained neural network to produce an output. Subsequently, the obtained output is compared with the desired output via a loss function. Then, gradient of the loss function is taken to be utilized in gradient descent algorithm to update the values of the physical features in a manner to minimize the loss function. This process is repeated many times that might result in convergence of the loss function to it's local or global minimum. The inverse design is an optimization problem that needs an optimizer to update values of the physical features through gradient descent. The optimizer utilized in this paper is known as Adam optimizer \cite{kingma2014adam} which is one of the most well-known and well-developed optimizer used in neural networks community.

\begin{figure*}
\centering
\includegraphics[width=0.9\textwidth]{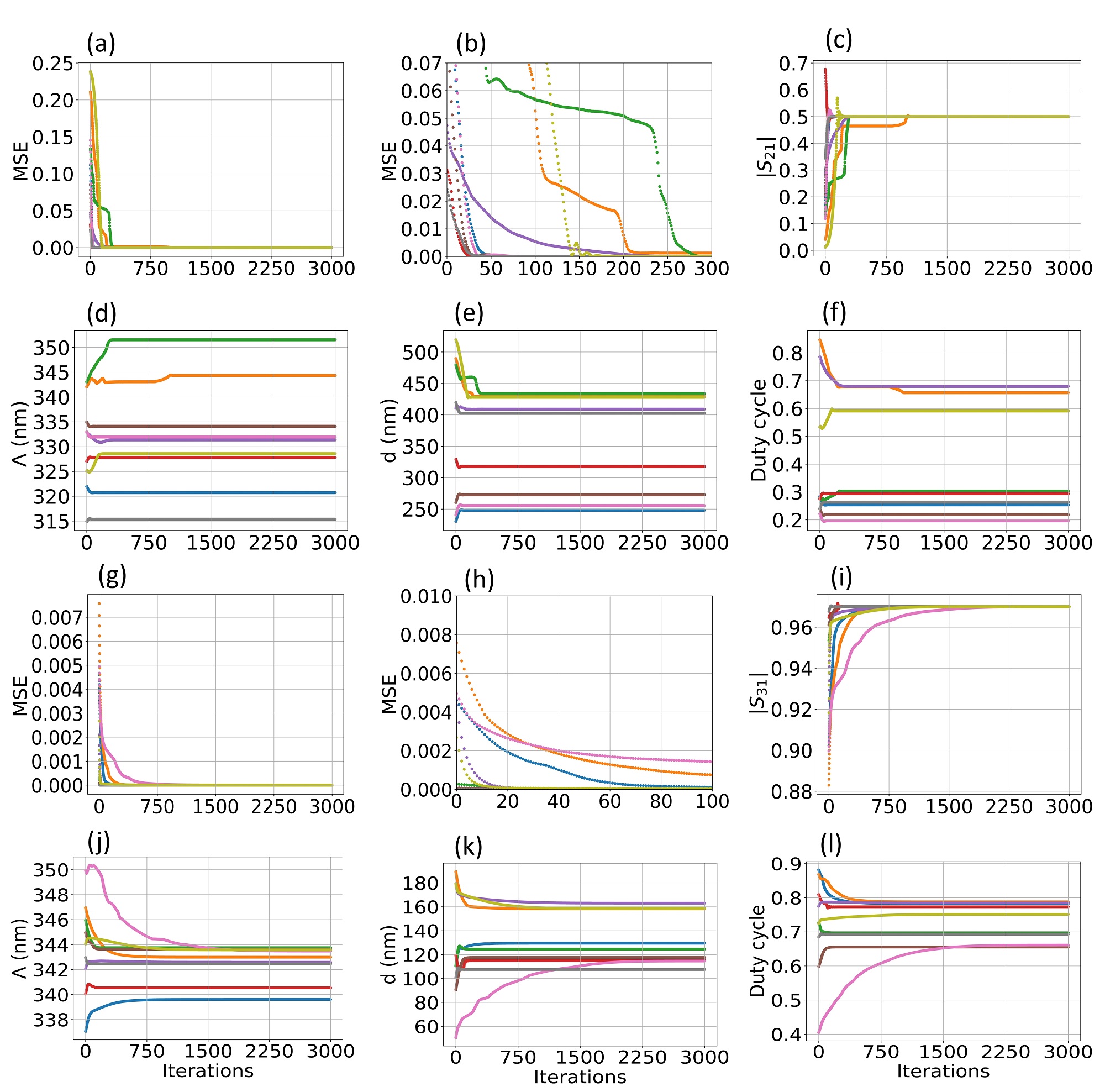}
\caption{(a)-(f) depict the variations of different parameters during the operation of the inverse design algorithm. The inverse design is performed 9 times to fulfill the desired value of $0.5$ for $|S_{21}|$.  As it is evident from the figures, the iteration number is chosen to be 3000. The optimization process starts with randomly selecting physical features of the grating. In the figures (a) and (b), MSE stands for mean-squared error. The figure (b) illustrates the zoomed version of the figure (a). In figures (d) and (e), \(\Lambda\) and $\mathbf{d}$ represent the grating period and the corrugation depth respectively. Figures (g)-(l) show the similar investigation as figures (a)-(f), however the inverse design is performed for the scattering parameter $|S_{31}|$ for the desired value equal to $0.97$.}
\label{evolution_s21_s31}
\end{figure*}

If the inverse design loss function and the function representing the trained neural network are denoted by the symbols $\textbf{L}$ and $\textbf{J}$, respectively, then the mathematical form of the loss function and update equations are as follows:

\begin{equation*}
\textbf{L} =\frac{1}{n}\ \sum_{1}^{n}\Bigl(\textbf{J}(\Lambda_{m},d_{m},t_{m})-\textbf{J}_{d}\Bigl)^2= 
\label{loss_function_inverse_nn_one_output}
\end{equation*}

\begin{equation}
(|S_{ij}^{m}|-|\hat{S}_{ij}|)^2 
\label{}
\end{equation}

\begin{equation}
\Lambda_{m+1} =\Lambda_{m} - \alpha \frac{\partial \, \textbf{L}}{\partial\, \Lambda_{m}} 
\label{update_eq_period_nn_one_output}
\end{equation}

\begin{equation}
d_{m+1} =d_{m} - \alpha \frac{\partial \, \textbf{L}}{\partial\, d_{m}} 
\label{update_eq_depth_nn_one_output}
\end{equation}

\begin{equation}
t_{m+1} =t_{m} - \alpha \frac{\partial \, \textbf{L}}{\partial\,t_{m}} 
\label{update_eq_duty_cycle_nn_one_output}
\end{equation}
where the parameter "$m$" denotes the iteration number in the gradient descent process. The terms $\textbf{J}_{d}$ and $|\hat{S}_{ij}|$ represent the target value. The symbol $|S_{ij}^{m}|$ denotes the trained network output. The subscript "$ij$" is relevant to the scattering parameter under investigation. The parameter "$\alpha$", in the update equations, represents the learning rate of the optimizer. The parameter "$n$" denotes the number of outputs of the trained network, and the summation is taking average. In this project, we have a single-output neural network. 

Figures~4(a-f) depict the results of nine-times applying the inverse design algorithm in the case that the desired value of $|S_{21}|$ is equal to $0.5$. Figure~4(a-b) illustrates evolution of the loss function to its minimum value. For all nine cases, the end values of the mean-squared error function are smaller than $10^{-10}$. Figures~4(d-f) are illustrating evolution of the values of the grating periods, the corrugation depths and the duty cycles from their random initial values to their end values. We decide that the repetition loop in the algorithm iterates 3000 times. Figure~4(c) illustrates evolution of $|S_{21}|$ for the nine cases. In a designer view, it might be interesting that we could have several different cases are capable of producing value $|S_{21}|$ equal to 0.5. This access to multiple physical features leading to the same $|S_{21}|$ might provide easier situation for manufacturing the grating structure.

The same type of investigation is done for $|S_{31}|$ in the case that the desired value is equal to 0.97. Figures~4(g-l) illustrate evolution of the physical features and the loss function for $|S_{31}|$.

\Table{\label{table1_S21_0.5}Inverse design for $|S_{21}|$=0.5. The last column contains values of $|S_{21}|$ obtained from electromagnetic simulation.} 
\br
Period ($nm$)&Depth ($nm$)& Duty& $|S_{21}|$ \\ 
\mr
320.72&248.19&0.25&0.5241\\
344.34&429.39&0.65&0.5086\\
351.53&433.8&0.3&0.5003\\
327.8&317.82&0.29&0.5223\\
331.35&408.97&0.67&0.4505\\
334.1&272.74&0.218&0.5083\\
331.94&255.61&0.19&0.5613\\
315.37&402.25&0.26&0.5663\\
328.58&428.05&0.59&0.4700\\
\br
\endTable

\Table{\label{table2_S31_0.97}Inverse design for $|S_{31}|$=0.97. The last column contains values of $|S_{31}|$ obtained from electromagnetic simulation.} 
\br
Period ($nm$)&Depth ($nm$)& Duty& $|S_{31}|$ \\ 
\mr
339.6&129.53&0.78&0.9696\\
342.98&158.31&0.78&0.9669\\
343.74&124.6&0.69&0.9666\\
340.53&115.02&0.77&0.9671\\
342.58&163&0.78&0.9693\\
343.62&117.59&0.65&0.9631\\
343.48&114.65&0.66&0.9686\\
342.45&107.56&0.69&0.9706\\
343.57&158.68&0.75&0.9636\\
\br
\endTable

To check whether the results of the inverse design are true, we need to test the achieved physical features by applying them in the software package and calculating the scattering parameters. In tables~\ref{table1_S21_0.5} and~\ref{table2_S31_0.97}, the values of $|S_{21}|$ and $|S_{31}|$ calculated by the software are shown respectively. The values of the physical features used in the simulation are the same as the ones shown in the figure~\ref{evolution_s21_s31} and are listed in the tables~\ref{table1_S21_0.5} and~\ref{table2_S31_0.97}. In the table ~\ref{table2_S31_0.97}, the difference between the desired value of $|S_{31}|$ which is equal to 0.97 and the simulated results is tiny. The reason is that the neural network model discussed in the section \ref{section:grating_waveguide_simulation}, works properly in mapping the input data set to the output. The quantitative approach to evaluate how good our neural network models are is to calculate the mean-squared error, the $R^2$ score and the explained-variance score of the predicted values and the actual values of the relevant scattering parameter. These evaluation numbers are arranged in the table~\ref{evaluation_table}. The $R^2$ score is a statistical measure that explains how much the independent variables are capable of explaining the variation in the dependent variable(s) in a regression model~\cite{weiming2019mastering}. In our case, the independent variables are the physical features of the grating and the dependent variable is the scattering parameter at the output of the neural network. The ideal value for $R^2$ score is equal to one, meaning that there is no error in the prediction ability of the model. The $R^2$ score value between 0.9 to 1 might indicate a good model. For our neural networks, the $R^2$ score is calculated for a test set that consists of 10\% of the whole data set. The prediction quality of the model for $|S_{31}|$ is also evident from the figure~3(l), where the predictions represented by the scatter points are very close to the red line illustrating an ideal model. The explained-variance score is another statistical measure used in regression analysis. The measure explains how dispersed the errors in the regression model. The best value of the explained-variance score for a model is equal to one. An example for the explained-variance score is that the measure for the model with epoch number equal to 5, depicted in the figure~3(i), is equal to 0.54 and for the model with epoch number 10000, illustrated in the figure~3(l), is equal to 0.99.

\Table{\label{evaluation_table}Statistical measures for the neural network models. MAE, MSE and EVS stand for mean-absolute error, mean-squared error and explained-variance score respectively. NN stands for neural network.}
\br
NN&MAE&MSE&$R^2$ score&EVS\\ 
\mr
For$|S_{21}|$ &0.0150&0.0010&0.9605&0.9605\\
For$|S_{31}|$ &0.0087&0.0002&0.9915&0.9915\\
\br
\endTable

In the case of the model for $|S_{21}|$, the mean-squared error and the explained-variance score is a little bit smaller than the case of $|S_{31}|$. This difference could be noted in the figures~3(h)-(l) where the spread of the point for $|S_{21}|$ model is fairly larger than for $|S_{31}|$ model. It is crucial to note that no model is prefect and the statistical measures should be evaluated in the context of data that is modeled.

\section{Conclusion}
\label{section:Conclusion}
Due to limitations of our knowledge and intuition of complex phenomena in photonic devices, inverse design of photonic components is becoming an important step of the whole process of creating those devices. Among the useful methods of inverse design and modeling, deep neural networks have strong position. In this paper we diligently tried to show that strength in modeling and inverse design of a waveguide grating which is an integral part many types of photonic devices. In the first step of this study, we collect data by simulating the grating structure. We sweep over three physical features of the grating i.e. the period, the depth of corrugations and the duty cycle. By changing those physical features, we collect the scattering parameters $|S_{11}|$, $|S_{21}|$, $|S_{31}|$. The number of data we collect is equal to 50545. For all cases, the input mode is the first TE mode of the waveguide and we investigate conversion of the $1^{st}$, the $2^{nd}$ and the $3^{rd}$ modes. It is possible to collect more data particularly by sweeping wider range of the period. The other two physical features of the grating have limitations to change, because the corrugation depth does not exceed 520 nm and the duty cycle can be altered in the range of 0 to 1.  

The second step is to design neural networks to map the physical features to each scattering parameter $|S_{11}|$, $|S_{21}|$, or $|S_{31}|$ separately. The neural networks play the role of a function with physical features as its inputs and the scattering parameter as its output. Having wider range of data is useful in this case because it helps increase the exactness of the neural network model and decrease the residual error.   

The third step is taking advantage of the designed neural networks in inverse designing of the grating. We define a criterion for the maximum value of the loss function in process of the inverse design. After obtaining the physical features that fulfil the desired value of the scattering parameter, it is important to test the feature in an electromagnetic simulation software to check weather the achieved features lead the same desired scattering parameter or not.

In general, the neural network approach works nicely in solving the problems like grating waveguide where, there is no neat formula to describe the behavior of the structure. However, maybe the main hardship relevant to using this approach is the data collection, which could be time-consuming and/or computationally demanding. An important positive view on the approach is that when the required data is collected, then the benefit from the efforts is long term, because  we have capability of establishing a fairly accurate function via neural networks that could be utilized in the future applications.

V.G. acknowledges support from Research Foundation Flanders under grant numbers G032822N and G0K9322N.

\section*{References}
\bibliographystyle{abbrv}
\bibliography{bibliography.bib}

\begin{thebibliography}{10}

\bibitem{alferness2013guided}
R.~Alferness, W.~Burns, J.~Donelly, I.~Kaminow, H.~Kogelnik, F.~Leonberger,
  A.~Milton, T.~Tamir, and R.~Tucker.
\newblock {\em Guided-wave optoelectronics}, volume~26.
\newblock Springer Science \& Business Media, 2013.

\bibitem{an2022deep}
S.~An, B.~Zheng, M.~Y. Shalaginov, H.~Tang, H.~Li, L.~Zhou, Y.~Dong,
  M.~Haerinia, A.~M. Agarwal, C.~Rivero-Baleine, et~al.
\newblock Deep convolutional neural networks to predict mutual coupling effects
  in metasurfaces.
\newblock {\em Advanced Optical Materials}, 10(3):2102113, 2022.

\bibitem{Dropout_layer}
J.~Brownlee.
\newblock Dropout regularization in deep learning models with keras,.

\bibitem{campbell2019review}
S.~D. Campbell, D.~Sell, R.~P. Jenkins, E.~B. Whiting, J.~A. Fan, and D.~H.
  Werner.
\newblock Review of numerical optimization techniques for meta-device design.
\newblock {\em Optical Materials Express}, 9(4):1842--1863, 2019.

\bibitem{carroll1998distributed}
J.~E. Carroll, J.~Whiteaway, D.~Plumb, and R.~Plumb.
\newblock {\em Distributed feedback semiconductor lasers}, volume~10.
\newblock IET, 1998.

\bibitem{christiansen2021inverse}
R.~E. Christiansen and O.~Sigmund.
\newblock Inverse design in photonics by topology optimization: tutorial.
\newblock {\em JOSA B}, 38(2):496--509, 2021.

\bibitem{chung2020tunable}
H.~Chung and O.~D. Miller.
\newblock Tunable metasurface inverse design for 80\% switching efficiencies
  and 144 angular deflection.
\newblock {\em Acs Photonics}, 7(8):2236--2243, 2020.

\bibitem{ginis2023resonators}
V.~Ginis, I.-C. Benea-Chelmus, J.~Lu, M.~Piccardo, and F.~Capasso.
\newblock Resonators with tailored optical path by cascaded-mode conversions.
\newblock {\em Nature Communications}, 14(1):495, 2023.

\bibitem{ginis2020remote}
V.~Ginis, M.~Piccardo, M.~Tamagnone, J.~Lu, M.~Qiu, S.~Kheifets, and
  F.~Capasso.
\newblock Remote structuring of near-field landscapes.
\newblock {\em Science}, 369(6502):436--440, 2020.

\bibitem{haupt2007genetic}
R.~L. Haupt and D.~H. Werner.
\newblock {\em Genetic algorithms in electromagnetics}.
\newblock John Wiley \& Sons, 2007.

\bibitem{jafar2018adaptive}
S.~Jafar-Zanjani, S.~Inampudi, and H.~Mosallaei.
\newblock Adaptive genetic algorithm for optical metasurfaces design.
\newblock {\em Scientific reports}, 8(1):1--16, 2018.

\bibitem{jensen2011topology}
J.~S. Jensen and O.~Sigmund.
\newblock Topology optimization for nano-photonics.
\newblock {\em Laser \& Photonics Reviews}, 5(2):308--321, 2011.

\bibitem{jiang2021deep}
J.~Jiang, M.~Chen, and J.~A. Fan.
\newblock Deep neural networks for the evaluation and design of photonic
  devices.
\newblock {\em Nature Reviews Materials}, 6(8):679--700, 2021.

\bibitem{jiang2019free}
J.~Jiang, D.~Sell, S.~Hoyer, J.~Hickey, J.~Yang, and J.~A. Fan.
\newblock Free-form diffractive metagrating design based on generative
  adversarial networks.
\newblock {\em ACS nano}, 13(8):8872--8878, 2019.

\bibitem{kingma2014adam}
D.~P. Kingma and J.~Ba.
\newblock Adam: A method for stochastic optimization.
\newblock {\em arXiv preprint arXiv:1412.6980}, 2014.

\bibitem{lenaerts2021artificial}
J.~Lenaerts, H.~Pinson, and V.~Ginis.
\newblock Artificial neural networks for inverse design of resonant
  nanophotonic components with oscillatory loss landscapes.
\newblock {\em Nanophotonics}, 10(1):385--392, 2021.

\bibitem{li2019inverse}
J.~Li, L.~Bao, S.~Jiang, Q.~Guo, D.~Xu, B.~Xiong, G.~Zhang, and F.~Yi.
\newblock Inverse design of multifunctional plasmonic metamaterial absorbers
  for infrared polarimetric imaging.
\newblock {\em Optics express}, 27(6):8375--8386, 2019.

\bibitem{liu2018generative}
Z.~Liu, D.~Zhu, S.~P. Rodrigues, K.-T. Lee, and W.~Cai.
\newblock Generative model for the inverse design of metasurfaces.
\newblock {\em Nano letters}, 18(10):6570--6576, 2018.

\bibitem{mao2021inverse}
S.~Mao, L.~Cheng, C.~Zhao, F.~N. Khan, Q.~Li, and H.~Fu.
\newblock Inverse design for silicon photonics: From iterative optimization
  algorithms to deep neural networks.
\newblock {\em Applied Sciences}, 11(9):3822, 2021.

\bibitem{mikki2008particle}
S.~M. Mikki and A.~A. Kishk.
\newblock Particle swarm optimization: A physics-based approach.
\newblock {\em Synthesis lectures on computational electromagnetics},
  3(1):1--103, 2008.

\bibitem{molesky2018inverse}
S.~Molesky, Z.~Lin, A.~Y. Piggott, W.~Jin, J.~Vuckovi{\'c}, and A.~W.
  Rodriguez.
\newblock Inverse design in nanophotonics.
\newblock {\em Nature Photonics}, 12(11):659--670, 2018.

\bibitem{peurifoy2018nanophotonic}
J.~Peurifoy, Y.~Shen, L.~Jing, Y.~Yang, F.~Cano-Renteria, B.~G. DeLacy, J.~D.
  Joannopoulos, M.~Tegmark, and M.~Solja{\v{c}}i{\'c}.
\newblock Nanophotonic particle simulation and inverse design using artificial
  neural networks.
\newblock {\em Science advances}, 4(6):eaar4206, 2018.

\bibitem{sajedian2019finding}
I.~Sajedian, J.~Kim, and J.~Rho.
\newblock Finding the optical properties of plasmonic structures by image
  processing using a combination of convolutional neural networks and recurrent
  neural networks.
\newblock {\em Microsystems \& nanoengineering}, 5(1):1--8, 2019.

\bibitem{shi2017optimization}
Y.~Shi, W.~Li, A.~Raman, and S.~Fan.
\newblock Optimization of multilayer optical films with a memetic algorithm and
  mixed integer programming.
\newblock {\em Acs Photonics}, 5(3):684--691, 2017.

\bibitem{so2020deep}
S.~So, T.~Badloe, J.~Noh, J.~Bravo-Abad, and J.~Rho.
\newblock Deep learning enabled inverse design in nanophotonics.
\newblock {\em Nanophotonics}, 9(5):1041--1057, 2020.

\bibitem{so2019designing}
S.~So and J.~Rho.
\newblock Designing nanophotonic structures using conditional deep
  convolutional generative adversarial networks.
\newblock {\em Nanophotonics}, 8(7):1255--1261, 2019.

\bibitem{venghaus2006wavelength}
H.~Venghaus.
\newblock {\em Wavelength filters in fibre optics}, volume 123.
\newblock springer, 2006.

\bibitem{weiming2019mastering}
J.~M. Weiming.
\newblock {\em Mastering Python for Finance: Implement advanced
  state-of-the-art financial statistical applications using Python}.
\newblock Packt Publishing Ltd, 2019.

\bibitem{wiecha2019design}
P.~R. Wiecha, C.~Majorel, C.~Girard, A.~Cuche, V.~Paillard, O.~L. Muskens, and
  A.~Arbouet.
\newblock Design of plasmonic directional antennas via evolutionary
  optimization.
\newblock {\em Optics express}, 27(20):29069--29081, 2019.

\bibitem{yariv1977periodic}
A.~Yariv and M.~Nakamura.
\newblock Periodic structures for integrated optics.
\newblock {\em IEEE journal of quantum electronics}, 13(4):233--253, 1977.

\bibitem{yariv2007photonics}
A.~Yariv and P.~Yeh.
\newblock {\em Photonics: optical electronics in modern communications}.
\newblock Oxford university press, 2007.

\end{thebibliography}

\end{document}